\documentclass[12pt]{article}

\catcode`\@=11
\@addtoreset{equation}{section}

\global\arraycolsep=2pt
\usepackage{mathrsfs,amsbsy,amssymb,latexsym,amsfonts,amsmath,cite}
\usepackage{graphicx,color}

\allowdisplaybreaks

\begin{document}

\begin{flushright}
\parbox{4.2cm}
{KUNS-2350}
\end{flushright}

\vspace*{2cm}

\begin{center}
{\Large \bf Hybrid classical integrability \\ 
in squashed sigma models}
\vspace*{2cm}\\
{\large Io Kawaguchi\footnote{E-mail:~io@gauge.scphys.kyoto-u.ac.jp} 
and 
Kentaroh Yoshida\footnote{E-mail:~kyoshida@gauge.scphys.kyoto-u.ac.jp} 
}
\end{center}

\vspace*{1cm}
\begin{center}
{\it Department of Physics, Kyoto University \\ 
Kyoto 606-8502, Japan} 
\end{center}

\vspace{1cm}
\begin{abstract}
We show that $SU(2)_{\rm L}$ Yangian and $q$-deformed $SU(2)_{\rm R}$ symmetries are realized 
in a two-dimensional sigma model defined on a three-dimensional squashed sphere. 
These symmetries enable us to develop the two descriptions to describe its classical dynamics, 
1) rational and 2) trigonometric descriptions.  The former 1) is based on the $SU(2)_{\rm L}$ symmetry and 
the latter 2) comes from the broken $SU(2)_{\rm R}$ symmetry. Each of the Lax pairs constructed in both ways 
leads to the same equations of motion. 
The two descriptions are related one another through a non-local map.  
\end{abstract}

\newpage

\section{Introduction}

The notion of integrability is of significance in theoretical and mathematical physics. 
It enables us to study physical quantities non-perturbatively and often prove 
strong-weak dualities exactly as in the case of sine-Gordon and massive Thirring models \cite{Coleman}. 
Similarly, integrability would be an important building block toward the proof of AdS/CFT \cite{Maldacena} 
(For an overview, see \cite{review}).  
In this direction the symmetric coset structure of AdS spaces and spheres would play an important role \cite{BPR}. 
A classification of symmetric cosets potentially applicable to AdS/CFT is performed in \cite{Zarembo}.

In applications of AdS/CFT to condensed matter physics,  
there is a motive to consider gravitational backgrounds, 
such as Schr$\ddot{\rm o}$dinger \cite{Son,BM} and Lifshitz \cite{Kachru} spacetimes, 
represented by non-symmetric cosets \cite{SYY}.  
As other examples, anisotropic geometries like warped AdS spaces and squashed spheres also appear 
as gravity duals to field theories in the presence of a magnetic field \cite{Kraus}. 
In condensed matter physics a magnetic field is of importance to vary the system, and hence 
the anisotropic geometries are very interesting to study. 

In this letter we will focus upon the classical integrable structure of a two-dimensional sigma model defined 
on a three-dimensional squashed sphere. Since the squashed sphere is described as a non-symmetric coset, 
it is not so obvious in comparison to symmetric cases such as principal chiral models \cite{Luscher2}. 

The squashed sphere is described as a one-parameter deformation 
of round $S^3$ and the metric of squashed $S^3$ 
\begin{eqnarray}
ds^{2} 
&=&-\frac{L^{2}}{2}\left[
{\rm Tr}\left(J^{2}\right)
-2C\left({\rm Tr}\left[T^{3}J\right]\right)^{2} \right] 
\label{squashed}
\end{eqnarray}
is represented by the left-invariant one-form $J \equiv g^{-1}dg$ with an $SU(2)$ group element $g$.  
The $SU(2)$ generators $T^{a}~(a=1,2,3)$ satisfy  
\begin{eqnarray}
\left[T^{a},T^{b}\right] = \varepsilon^{ab}_{~~c}T^{c}\,, \quad 
{\rm Tr}\left(T^{a}T^{b}\right) = -\frac{1}{2}\delta^{ab}\,, \nonumber 
\end{eqnarray}
where $\varepsilon^{ab}_{~~c}$ is the totally antisymmetric tensor. 
The constant $C$ measures the deformation from $S^3$\,. 
When $C=0$\,, the metric (\ref{squashed}) describes the round $S^3$ with radius $L$\,. 
For $C \neq 0$\,, the $S^3$ isometry $SO(4)=SU(2)_{\rm L} \times SU(2)_{\rm R}$ is broken to 
$SU(2)_{\rm L} \times U(1)_{\rm R}$\,. 
The infinitesimal transformations under $SU(2)_{\rm L} \times U(1)_{\rm R}$ 
are given by 
\begin{eqnarray}
\delta^{L,a}g = \epsilon_L\,T^{a}g\,, \qquad \delta^{R,3}g = -\epsilon_R\,gT^{3}\,. 
\label{trans-1}
\end{eqnarray}

Let us consider a two-dimensional non-linear sigma model whose target space is the squashed sphere (\ref{squashed}). 
The action is given by 
\begin{eqnarray}
S =\!\! \int\!\!\!\!\int\!\!dtdx\Bigl[
{\rm Tr}\left(J_{\mu} J^{\mu}\right)
-2C\,{\rm Tr}\!\left(T^{3}J_{\mu}\right)\!{\rm Tr}\!\left(T^{3}J^{\mu}\right)
\Bigr]. 
\label{sigma}
\end{eqnarray}
The coordinates and metric of base space are $x^{\mu}=(t,x)$ and $\eta_{\mu\nu}={\rm diag}(-1,+1)$\,. 
Suppose that the value of $C$ is restricted to $C > -1$ 
so that the sign of kinetic term is not flipped. 
The action (\ref{sigma}) is invariant under (\ref{trans-1}). 

Note that the Virasoro and periodic boundary conditions are not imposed here. 
Instead, we impose the boundary condition that the variable $g(x)$ approaches a constant element rapidly 
as it goes to spatial infinity, 
\begin{eqnarray}
g(x) \to g_{\pm\infty}~:~\mbox{const.} \qquad (x \to \pm \infty)\,. \label{bc}
\end{eqnarray}
That is, $J_{\mu}(x)$ vanishes rapidly as $x\to \pm \infty$\,. 

The equations of motion are 
\begin{eqnarray}
\partial^{\mu}J_{\mu} - 2C {\rm Tr}(T^3\partial^{\mu}J_{\mu})T^3 - 2C\,{\rm Tr}(T^3J_{\mu})[J^{\mu},T^3] =0\,.
\label{eom}
\end{eqnarray}
Multiplying $T^3$ and taking the trace, we obtain the conservation law for $U(1)_{\rm R}$, 
\begin{eqnarray}
\partial^{\mu}{\rm Tr}(T^3 J_{\mu}) = 0\,. \label{U(1)R}
\label{u1conserve}
\end{eqnarray}
Then the expressions in (\ref{eom}) are simplified as 
\begin{eqnarray}
\partial^{\mu}J_{\mu} - 2C\,{\rm Tr}(T^3J_{\mu})[J^{\mu},T^3] =0\,. 
\label{eom2} 
\end{eqnarray}
We will show that the equations of motion (\ref{eom2}) are reproduced from the two descriptions, 
1) the rational description with $SU(2)_{\rm L}$ and 2) the trigonometric one with $U(1)_{\rm R}$\,.

\section{Rational description}

First let us consider a description based on the $SU(2)_{\rm L}$ symmetry.  
The $SU(2)_{\rm L}$ Noether current $j_{\mu}^L$ is given by 
\begin{eqnarray}
j_{\mu}^{L} = g J_{\mu} g^{-1}-2C{\rm Tr}\left(T^{3}J_{\mu}\right)gT^{3}g^{-1}\,. 
\label{SU(2)}
\end{eqnarray}
Then the conservation laws follow from (\ref{eom2}). 
The number of dynamical degrees of freedom in this system is just three.  
It agrees with that of the conserved charges for $SU(2)_{\rm L}$\,. Thus the equations of motion and 
the conservation laws of $SU(2)_{\rm L}$ are equivalent. The $U(1)_{\rm R}$ current is 
automatically conserved due to the conservation laws of $SU(2)_{\rm L}$\,. 

Although the current (\ref{SU(2)}) does not satisfy the flatness condition,  
it can be improved by adding a topological term so that it does. 
The improved current $\tilde{j}_{\mu}^L$ is given by
\begin{eqnarray}
\tilde{j}^{L}_{\mu} \equiv j_{\mu}^{L} 
- \sqrt{C}\,\epsilon_{\mu\nu}\partial^{\nu}\left(gT^{3}g^{-1}\right)\,, \label{improve}
\end{eqnarray}
and satisfies the flatness condition \cite{KY}:
\begin{eqnarray}
\epsilon^{\mu\nu}\left(\partial_{\mu}\tilde{j}^{L}_{\nu}
- \tilde{j}^{L}_{\mu}\tilde{j}^{L}_{\nu}\right)=0\,. \label{flat}
\end{eqnarray}
The anti-symmetric tensor $\epsilon_{\mu\nu}$ on the base space is normalized as $\epsilon_{tx}=+1$\,. 
The coefficient of the last term in (\ref{improve}) is fixed so that the flat condition (\ref{flat}) is satisfied. 

For the improved $SU(2)_{\rm L}$ current, the current algebra is deformed by the squashing parameter $C$ as follows: 
\begin{eqnarray}
\bigl\{\tilde{j}^{L,a}_{t}(x),\, \tilde{j}^{L,b}_{t}(y)\bigr\}_{\rm P} 
&=& \varepsilon^{ab}_{~~c}\, \tilde{j}^{L,c}_{t}(x)\,\delta(x-y)\,, \nonumber \\
\bigl\{\tilde{j}^{L,a}_{t}(x),\, \tilde{j}^{L,b}_{x}(y)\bigr\}_{\rm P} 
&=& \varepsilon^{ab}_{~~c}\, \tilde{j}^{L,c}_{x}(x)\,\delta(x-y) 
 +(1+C)\delta^{ab}\,\partial_{x}\delta(x-y)\,, \nonumber \\
\bigl\{\tilde{j}^{L,a}_{x}(x),\, \tilde{j}^{L,b}_{x}(y)\bigr\}_{\rm P} 
&=& -C\,\varepsilon^{ab}_{~~c}\, \tilde{j}^{L,c}_{t}(x)\,\delta(x-y) \nonumber 
\end{eqnarray}
Here we have used the vector index notation with $\tilde{j}_{\mu}^{L,a} \equiv -2{\rm Tr}(T^a \tilde{j}_{\mu}^{L})$\,. 
Due to the improvement, 
an infinite number of conserved charges can be constructed, for example, by following \cite{BIZZ}.
The first two of them are 
\begin{eqnarray}
Q^{L,a}_{(0)}&=&\int\!\!dx\, \tilde{j}^{L,a}_{t}(x)\,, \nonumber \\
Q^{L,a}_{(1)}&=& -\int\!\!dx\, \tilde{j}^{L,a}_{x}(x) 
 + \frac{1}{4}\int\!\!\!\!\int\!\!dxdy\,\epsilon(x-y)\varepsilon^{a}_{~bc}\, \tilde{j}^{L,b}_{t}(x)\tilde{j}^{L,c}_{t}(y)\,. 
\nonumber 
\end{eqnarray} 
Here $\epsilon(x-y) \equiv \theta (x-y) - \theta(y-x)$ and $\theta(x)$ is a step function. 
Although the current algebra is deformed, the Yangian algebra \cite{Drinfeld} is still realized and 
the Serre relations are also satisfied \cite{KY}. This is the case even after adding the Wess-Zumino term \cite{KOY}, though 
the current algebra becomes much more complicated. 

It is a turn to construct a Lax pair. 
With the improved $SU(2)_{\rm L}$ current, it can be constructed as a linear combination,  
\begin{eqnarray}
L_{\mu}^L(x;\lambda)=\frac{\lambda}{\lambda^{2}-1}\left[\epsilon_{\mu\nu} \tilde{j}^{\nu\, L}(x) 
+ \lambda \tilde{j}^{L}_{\mu}(x)\right]\,, 
\label{Lax}
\end{eqnarray}
where $\lambda$ is a spectral parameter. 
Then the commutation relation
\begin{eqnarray}
\left[\partial_{t}-L_{t}^L(\lambda),
\partial_{x}- L_{x}^L(\lambda)\right] = 0 
\label{flat-Lax}
\end{eqnarray}
leads to the whole equations of motion (\ref{eom2})\,. 

With (\ref{Lax}), the monodromy matrix $U^L(\lambda)$ is defined as 
\begin{eqnarray}
U^L(\lambda) \equiv {\rm P}\exp{\left[\int^{\infty}_{-\infty}\!\!dx\,L_{x}^L(x;\lambda)\right]}\,. \nonumber
\end{eqnarray}
The symbol P means the path ordering. 
Due to (\ref{flat-Lax}), $U^L(\lambda)$ is conserved,  
\begin{eqnarray}
\frac{d}{dt}U^L(\lambda)=0\,. \nonumber
\end{eqnarray}
Thus one can obtain an infinite number of conserved charges by expanding $U^L(\lambda)$ around a fixed value of $\lambda$\,. 
The expressions of the charges depend on expansion points.  
The expansion around $\lambda=0$ leads to an infinite number of the non-local charges constructed in \cite{KY}.  
When it is expanded around $\lambda=\pm 1$\,, an infinite number of commuting local charges 
which ensure the classical integrability in the sense of Liouville. 

The classical $r$-matrix is derived by evaluating the Poisson bracket of the monodromy matrices.  
Following the prescription in \cite{Duncan}, it is evaluated as 
\begin{eqnarray}
\left\{U^L(\lambda)^{i}_{~j}\,,U^L(\mu)^{k}_{~l}\right\}_{\rm P}&=&\left[r^L(\lambda,\mu)\,, U^L(\lambda)\otimes U^L(\mu)\right]^{ik}_{jl}\,, 
\nonumber 
\end{eqnarray}
and the classical $r$-matrix is 
\begin{eqnarray}
r^L(\lambda,\mu)^{ik}_{jl}&=&\frac{\lambda\mu}{\lambda-\mu}\delta_{ab}(T^{a})^{i}_{~j}(T^{b})^{k}_{~l}\,. \nonumber
\end{eqnarray}
Note that the resulting $r$-matrix does not contain $C$ and is of the familiar rational type. 
Thus it satisfies the classical Yang-Baxter equation as a matter of course.

\section{Trigonometric description}

Next we shall consider another description based on the broken $SU(2)_{\rm R}$ symmetry. 
We first show that the broken $SU(2)_{\rm R}$ symmetry is enhanced to a $q$-deformed $SU(2)_{\rm R}$ 
symmetry. 

Recall that the $U(1)_{\rm R}$ current is given by 
\begin{eqnarray}
j^{R,3}_{\mu}=-2(1+C){\rm Tr}\left(T^{3}J_{\mu}\right)\,.
\label{norm}
\end{eqnarray}
The normalization is taken for later convenience. 

Now let us consider the following currents, 
\begin{eqnarray}
j^{R,\pm}_{\mu} &=& 
-2\,{\rm e}^{\gamma\chi}
\left(\eta_{\mu\nu} 
\pm i\sqrt{C}\epsilon_{\mu\nu}\right){\rm Tr}\left(T^{\pm}J^{\nu}\right)\,,
\label{non-local} 
\end{eqnarray}
where \[
\gamma = \frac{\sqrt{C}}{1+C}\,, \qquad T^{\pm} \equiv \frac{1}{\sqrt{2}}\left(T^{1}\pm i T^{2}\right)\,.
\]  
The field $\chi(x)$ contained in (\ref{non-local}) is given by 
\begin{eqnarray}
\chi(x) &=& \frac{1}{2}\int\!\!dy\,\epsilon(x-y)\,j^{R,3}_{t}(y)
\end{eqnarray}
and non-local. 
Thus the currents in (\ref{non-local}) are also non-local\footnote{Note that a non-local symmetry concerning $SU(2)_{\rm R}$ 
is discussed also in \cite{ORU} from a T-duality argument. However, the one discussed here 
is different from this. A modification is done motivated by \cite{Bernard}. }.  
To show the conservation laws of non-local currents in (\ref{non-local})\,, it is necessary to use the boundary condition (\ref{bc}) and the identities,  
\begin{eqnarray}
\partial^{\mu}\chi = \epsilon^{\mu\nu}j_{\nu}^{R,3}\,, \qquad 
\epsilon_{\mu\nu}\left[\pm i{\rm Tr}(T^{\pm}\partial^{\mu}J^{\nu}) 
-2 {\rm Tr}(T^{\pm}J^{\mu}) {\rm Tr}(T^{3}J^{\nu})\right] = 0 \,.
\nonumber 
\end{eqnarray}

The Poisson brackets of $j^{R,\pm}_{t}$ and $j^{R,3}_{t}$ are 
\begin{eqnarray}
\left\{j^{R,\pm}_{t}(x),\,j^{R,\mp}_{t}(y)\right\}_{\rm P}&=& \pm i\,{\rm e}^{2\gamma\,\chi(x)}\,
j^{R,3}_{t}(x)\delta(x-y) \nonumber \\
&=&\pm \,\frac{i}{2\gamma}\partial_{x}
\left[{\rm e}^{2\gamma\,\chi(x)}\right]\delta(x-y)\,, \nonumber \\
\left\{j^{R,\pm}_{t}(x),\,j^{R,\pm}_{t}(y)\right\}_{\rm P} 
&=& \pm i\, \gamma\,\epsilon(x-y)\,j_{t}^{R,\pm}(x)j_{t}^{R,\pm}(y)\,, \nonumber  \\ 
\left\{j^{R,3}_{t}(x),\,j^{R,\pm}_{t}(y)\right\}_{\rm P} 
&=& \pm i\,j^{R,\pm}_{t}(x)\delta(x-y)\,. \nonumber 
\end{eqnarray}
The conserved charges are constructed as
\begin{eqnarray}
Q^{R,\pm} = \int\!\!dx\,j^{R,\pm}_{t}(x)\,, \qquad Q^{R,3} = \int\!\!dx\,j^{R,3}_{t}(x)\,. \nonumber 
\end{eqnarray}
Then the transformation laws generated by $Q^{R,\pm}$ are 
\begin{eqnarray}
\delta^{R,\pm}g 
= \{Q^{R,\pm},g\}_{\rm P} = g\left[- T^{\pm} {\rm e}^{\gamma\,\chi} 
+ \gamma T^{3}\xi^{\pm}
\right]\,, \label{non-local trans}
\end{eqnarray}
where $\xi^{\pm}$ are new non-local fields given by  
\begin{eqnarray}
\xi^{\pm}(x) \equiv \frac{1}{2}\int\!\!dy\,\epsilon(x-y)\,j^{R,\pm}_{t}(y)\,. 
\nonumber 
\end{eqnarray}
It is now straightforward to check the invariance of the equations of motion (\ref{eom2}) directly 
and thus the transformation laws (\ref{non-local trans}) 
give rise to an ``on-shell'' symmetry. 
When $C=0$\,, the transformation laws (\ref{non-local trans}) are reduced to the usual $SU(2)_{\rm R}$ ones. 

Note that the relations $\chi(\pm\infty)=\pm Q^{R,3}/2$ hold under the boundary condition (\ref{bc})\,. 
Thus the Poisson brackets of the charges lead to a $q$-deformed $SU(2)$ algebra \cite{Drinfeld,Jimbo}~:
\begin{eqnarray}
&& \left\{Q^{R,+},Q^{R,-}\right\}_{\rm P} =i\,\frac{q^{Q^{R,3}}-q^{-Q^{R,3}}}{q-q^{-1}}\,, \qquad 
q \equiv {\rm e}^{\gamma} = \exp\left(\frac{\sqrt{C}}{1+C}\right)\,,  
\nonumber \\ 
&& \left\{Q^{R,3},Q^{R,\pm}\right\}_{\rm P} = \pm i\, Q^{R,\pm}\,. \label{q-deformed}
\end{eqnarray}
Here we have rescaled $Q^{R,\pm}$ as 
\[
Q^{R,\pm} \longrightarrow \left(\frac{\gamma}{\sinh\gamma} \right)^{1/2}
\, Q^{R,\pm}\,.
\]
The normalization of (\ref{norm}) is fixed so that the expression of the second commutator in (\ref{q-deformed}) 
is obtained. 

It is worth noting the $C\to 0$ limit where round $S^3$ is reproduced 
and hence the $SU(2)_{\rm R}$ Yangian should be recovered. This is the case as we can see 
by expanding the non-local currents in (\ref{non-local}) in terms of $\sqrt{C}$ like 
\begin{eqnarray}
Q^{R,\pm} = Q^{R,\pm}_{(0)} \pm i \sqrt{C}\, Q^{R,\pm}_{(1)} + \cdots \,, 
\end{eqnarray}
where $Q^{R,\pm}_{(0)}$ and $Q^{R,\pm}_{(1)}$ are the $SU(2)_{\rm R}$ Yangian generators. 
The third component of the Yangian generators is supplemented from the Poisson bracket of the $+$ and $-$ components. 

On the other hand, when considering the $C \to \infty$ limit, 
${\rm Tr}(T^{\pm}J_{\mu})$ and ${\rm Tr}(T^3 J_{\mu})$ have to vanish for the finiteness of $Q^{R,3}$.  
This implies that a single element of $SU(2)$ is specified. In analogy with the XXZ model, 
the $C\to \infty$ limit resembles the Ising model limit. The fact that a single point is preferred 
would be analogous to that a ferromagnetic ground state is picked up in the Ising model. 

It is a turn to consider a Lax pair given by \cite{FR}
\begin{eqnarray}
L^{R}_x &=& -\sum_{a=1}^3\left[ 
w_a(\lambda - \alpha) S^a + w_a(\lambda + \alpha) \bar{S}^a 
\right] T^a\,, \nonumber \\ 
L^{R}_t &=& -\sum_{a=1}^3\left[ 
w_a(\lambda - \alpha) S^a - w_a(\lambda + \alpha) \bar{S}^a 
\right] T^a\,.  \label{Lax-R} 
\end{eqnarray}
$S^{a}$ and $\bar{S}^{a}$ are related to $J_{t}^a$ and 
$J_x^a$ as follows: 
\begin{eqnarray}
&& J_t^3 = (w_1(2\alpha)+w_3(2\alpha))(S^3 + \bar{S}^3)\,, \nonumber \\
&& J_x^3 = (w_1(2\alpha)+w_3(2\alpha))(S^3 - \bar{S}^3)\,,  \nonumber \\ 
&& J_t^{1,2} = \sqrt{2 w_1(2\alpha)(w_1(2\alpha)+w_3(2\alpha))}\,(S^{1,2} + \bar{S}^{1,2})\,, \nonumber \\ 
&& J_x^{1,2} = \sqrt{2 w_1(2\alpha)(w_1(2\alpha)+w_3(2\alpha))}\,(S^{1,2} - \bar{S}^{1,2})\,. \nonumber 
\end{eqnarray}
Here $\lambda$ is a spectral parameter and $w_a(\lambda)$ are defined as 
\begin{eqnarray}
w_1(\lambda) = w_2(\lambda) \equiv \frac{1}{\sinh\lambda}\,, \quad w_3(\lambda) \equiv \coth\lambda\,. \nonumber
\end{eqnarray}
The location of pole $\alpha$ is specified as 
\begin{eqnarray} 
C = \frac{w_1(2\alpha) - w_3(2\alpha)}{w_1(2\alpha) + w_3(2\alpha)} =  -\tanh^2\,\alpha\,. \nonumber
\end{eqnarray}
By definition, $\alpha$ can take a complex value, while $C$ must be real. 
Therefore $\alpha$ should be real or purely imaginary. When we take $\alpha = i\beta$~($\beta$:~real)\,, 
then $C = \tan^2\,\beta$\,. Then the range of $C$ is naturally restricted 
to the physical region $C \geq -1$\,. 
By rescaling $\lambda$ as $\lambda = \alpha \tilde{\lambda}$ and taking the $\alpha \to 0$ limit in (\ref{Lax-R}), 
the Lax pair of rational type for $SU(2)_{\rm R}$ is reproduced. 

The commutation relation
\begin{eqnarray}
\left[\partial_{t}+L^{R}_{t}(\lambda),\partial_{x}+L^{R}_{x}(\lambda)\right]=0 \nonumber
\end{eqnarray}
leads to the equations of motion (\ref{eom2}) with the help of the flatness of $J=g^{-1}dg$\,.
Then the monodromy matrix is defined as 
\begin{eqnarray}
U^{R}(\lambda)\equiv{\rm P}\exp\left[-\int^{\infty}_{-\infty}\!\!dx\,L^{R}_{x}(x;\lambda)\right]\, \nonumber
\end{eqnarray}
and it is conserved,
\begin{eqnarray}
\frac{d}{dt}U^R(\lambda)=0\,. \nonumber 
\end{eqnarray}
Following the prescription in \cite{Duncan}, the Poisson bracket of the monodromy matrices is 
evaluated as 
\begin{eqnarray}
\left\{U^{R}(\lambda)^{i}_{~j},U^{R}(\mu)^{k}_{~l}\right\}_{\rm P}
=\left[r^{R}(\lambda,\mu),U^{R}(\lambda)\otimes U^{R}(\mu)\right]^{ik}_{jl}\,. \nonumber 
\end{eqnarray}
The resulting classical $r$-matrix is given by 
\begin{eqnarray}
r^{R}(\lambda,\mu)^{ik}_{jl} &\equiv & 
\frac{1}{\sinh{(\lambda\!-\!\mu)}}\!\left[(T_{+})^{i}_{~j}(T_{-})^{k}_{~l}\!+\!(T_{-})^{i}_{~j}(T_{+})^{k}_{~l}
+\!\cosh{(\lambda\!-\!\mu)}(T_{3})^{i}_{~j}(T_{3})^{k}_{~l}\right] \nonumber 
\end{eqnarray}
and is of trigonometric type. This classical $r$-matrix also satisfies the Yang-Baxter equation.

Finally let us discuss the equivalence between the two descriptions. 
The current circumstance is quite similar to the Seiberg-Witten map \cite{SW}. On the one hand, 
The improvement term added in (\ref{improve}) may be regarded as a constant two-form flux. On the other hand,  
the existence of $q$-deformed $SU(2)_{\rm R}$ implies a ``quantum space'' such as a noncommutative space. 
In fact, $j_{\mu}^{R,a}$ can be expressed in terms of 
the improved $SU(2)_{\rm L}$ current $\tilde{j}_{\mu}^{L,a}$ like 
\begin{eqnarray}
j_{\mu}^{R,\pm} &=& -2\, {\rm e}^{\gamma\chi}\,{\rm Tr}(g^{-1}\tilde{j}_{\mu}^L gT^{\pm})\,,  \nonumber \\ 
j_{\mu}^{R,3} &=& -2\, {\rm Tr}(g^{-1}\tilde{j}_{\mu}^L g T^3)\,. \label{NL-map}
\end{eqnarray}
Thus the two descriptions discussed so far are not independent one another, as in the case of principal chiral models 
where both left and right currents are of rational type. It is remarkable that in the present case the trigonometric description is 
equivalent to the rational one through the non-local map (\ref{NL-map}).

\section{Discussions}

In this letter we have shown that $SU(2)_{\rm L}$ Yangian and $q$-deformed $SU(2)_{\rm R}$ symmetries are realized
in a two-dimensional sigma model defined on a three-dimensional squashed sphere. According to these hidden symmetries, 
we have presented the two descriptions,  
1) the rational description and 2) the trigonometric one. They are related one another via a non-local map 
and hence are equivalent. 
Recall that one may consider the Seiberg-Witten map in a field theory equipped with a magnetic field. 
On the other hand, warped AdS spaces, which are obtained from the squashed sphere through double Wick rotations,  
appear as gravity duals of condensed matter systems in the presence of a magnetic field. Therefore, the equivalence 
discussed here would be rather natural as a sigma model realization of Seiberg-Witten map in the field theory dual.  

The next question is what is the interpretation of this equivalence in gravitational theories. 
Warped AdS spaces appear also in the Kerr/CFT correspondence \cite{Kerr/CFT}. 
A three-dimensional slice of the near-horizon extreme Kerr geometry \cite{BH} 
is described as a warped AdS$_3$ space. It would be interesting to consider the role of 
$q$-deformed $SU(2)_{\rm R}$ in this direction. It may lead to a new source of entropy. 

Another issue is to construct the Bethe ansatz based on the $SU(2)_{\rm L}$ Yangian 
and $q$-deformed $SU(2)_{\rm R}$ symmetries. The speculated Bethe ansatz should be called ``hybrid'' Bethe ansatz 
which is composed of the S-matrices of XXX and XXZ models for the left and right, respectively. 
In fact, quantum solutions are already known \cite{quantum1,quantum2,quantum3}, 
though the classical integrable structure we revealed here has not been discussed there. 
It would be interesting to consider them in the context of AdS/CFT.

\subsection*{Acknowledgments}

We would like to thank D.~Orlando, R.~Sasaki and M.~Staudacher  
for illuminating discussions. The work of IK was supported by the Japan Society for the Promotion of Science (JSPS). 
The work of KY was supported by the scientific grants from the Ministry of Education, Culture, Sports, Science 
and Technology (MEXT) of Japan (No.\,22740160). 
This work was also supported in part by the Grant-in-Aid 
for the Global COE Program ``The Next Generation of Physics, Spun 
from Universality and Emergence'' from 
MEXT, Japan.

\end{document}